\documentclass[aps,prr,superscriptaddress,twocolumn
]{revtex4-1}
\usepackage{graphicx,amsmath,mathtools,amsfonts}
\usepackage{bm}
\usepackage[normalem]{ulem}
\usepackage[usenames, dvipsnames]{xcolor}

\usepackage{hyperref}
\hypersetup{colorlinks=true, linkcolor=blue, citecolor=blue, urlcolor=blue}

\begin{document}

\title{Single-layer T'-type nickelates: Ni$^{1+}$ is Ni$^{1+}$}

\author{F. Bernardini}
\affiliation{Dipartimento di Fisica, Universit\`a di Cagliari, IT-09042 Monserrato, Italy}

\author{A. Demourgues}
\affiliation{CNRS, Université Bordeaux, ICMCB, UMR 5026, 33600 Pessac, France}

\author{A. Cano}
\affiliation{
Univ. Grenoble Alpes, CNRS, Grenoble INP, Institut Néel, 25 Rue des Martyrs, 38042, Grenoble, France
}
\date{\today}

\begin{abstract}
The discovery of superconductivity in the infinite-layer nickelates has opened new perspectives in the context of quantum materials. We analyze, via first-principles calculations, the electronic properties of La$_2$NiO$_3$F ---the first single-layer T'-type nickelate--- and compare these properties with those of related nickelates and isostructural cuprates. We find that La$_2$NiO$_3$F is essentially a single-band system with a Fermi surface dominated by the Ni-3$d_{x^2-y^2}$ states with an exceptional 2D character. In addition, the hopping ratio is similar to that of the highest $T_c$ cuprates and there is a remarkable $e_g$ splitting together with a charge transfer energy of 3.6~eV. According to these descriptors, along with a comparison to Nd$_2$CuO$_4$, we thus indicate
single-layer T'-type nickelates of this class as very promising analogs of cuprate-like physics while keeping distinct Ni$^{1+}$ features.   
\end{abstract}

\maketitle

\section{Introduction}

The reduced Ruddlesen-Popper series $R_{n+1}$Ni$_n$O$_{2n+2}$ ($R = $ rare-earth element) provides a special class of nickelates. These compounds are of particular interest both as a chemistry challenge and as 
a new arena for quantum materials
\cite{review-nickelates20,norman20-p,pickett20-np}. 
Infinite-layer ($n=\infty$) nickelates are known since the early 1980s  \cite{crespin83,hayward99,hayward2003,crespin05}, while the synthesis of the $n =2$ and $3$ phases are more recent achievements \cite{poltavets1, poltavets2, Poltavets_3, Poltavets_4}. These phases display NiO$_2$ layers where the Ni atom formally features unusual $(1+{1/n})+$ oxidation states, and hence $d^{9-{1/ n}}$ electronic configurations tantalizingly similar to those in the superconducting cuprates for large $n$s \cite{anisimov99,pickett-prb04,zhang_nat_phys}.   
The latest discovery of 
unconventional superconductivity in the infinite-layer case has sparked a renewed interest on these systems \cite{hwang19a,ariando20,hwang20Pr-a,hwang20Pr-b,wen20a}. 
This long-sought finding has motivated a flurry of studies on the normal and superconducting-state properties of these systems \cite{review-nickelates20}, as well as the search for additional $d^9$ materials \cite{lorenzana19,arita20prb,cano20d}. 

\begin{figure}[b!]
    \includegraphics[width=.4\textwidth]{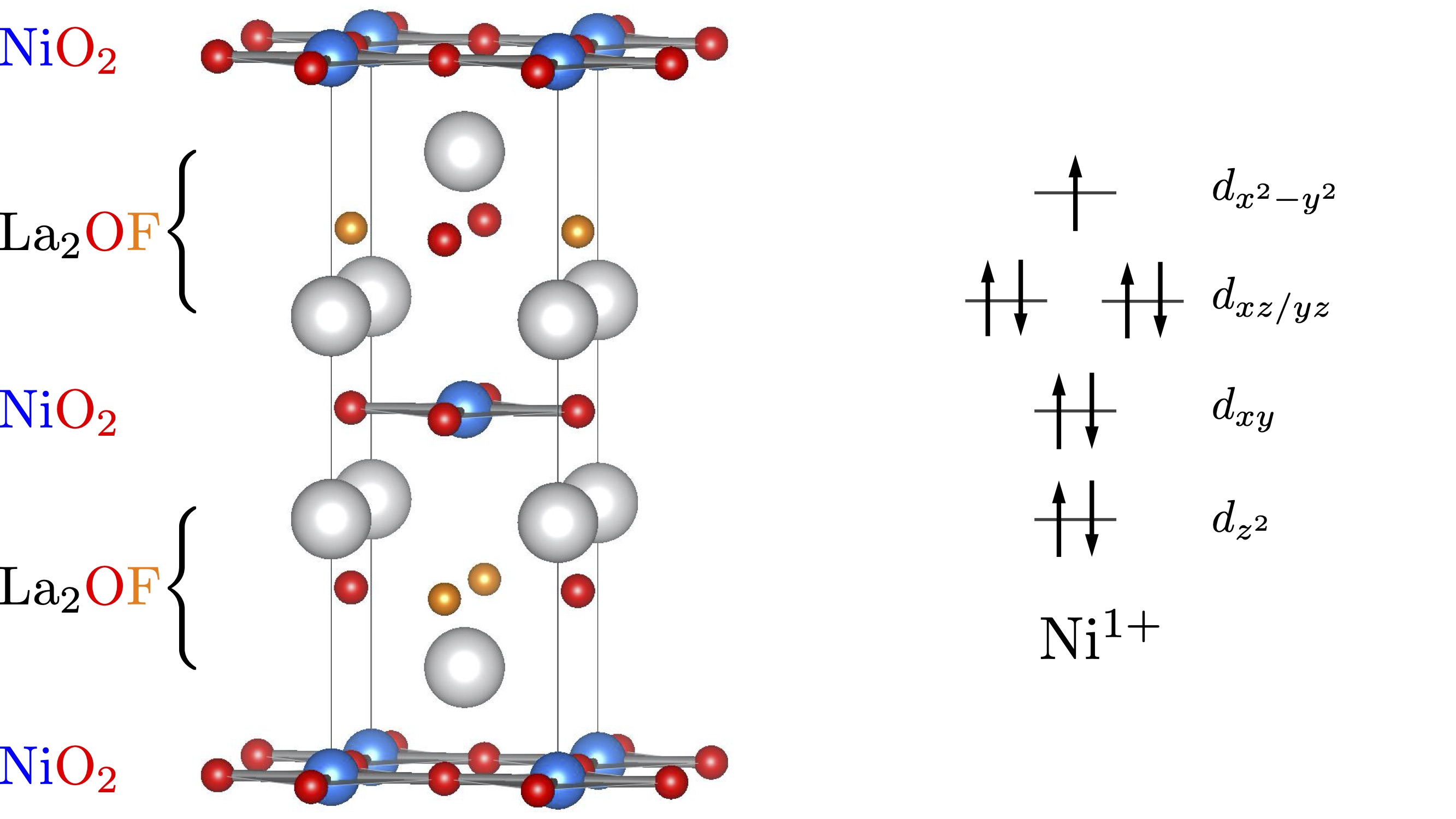}
    \caption{(a) Ball-and-stick model of the single-layer T'-type La$_2$NiO$_3$F where single NiO$_2$ layers are separated by La$_2$OF fluorite blocks (the conventional unit cell is indicated by the lines). (b) Nominal 3$d^9$ electronic configuration of the Ni$^{1+}$ atom in this single-layer T'-type class of nickelates.}
    \label{f:0}
\end{figure}

In this context, here we analyze the electronic structure of the novel La$_2$NiO$_3$F compound where the 3$d^9$ configuration of the Ni$^{1+}$ is formally realized as in the superconducting infinite-layer case. 
This system represents first $n=1$ T'-type nickelate ---isostructural to Nd$_2$CuO$_4$--- recently synthesized by Wissel {\it et al.} by means of an original two-step fluorination technique \cite{clemens20}. 
Thus, rather than the rock-salt spacer of the T-type K$_2$NiF$_4$ structures ({\it e.g.} La$_2$NiO$_4$ or La$_2$CuO$_4$), the T'-type La$_2$NiO$_3$F displays fluorite blocks separating the NiO$_2$ planes as illustrated in Fig. \ref{f:0}. 
This novel single-layer nickelate thus provides an additional juxtaposing angle between nickelates and cuprates.

We find that the potential analogy between nickelates and cuprate superconductors that has motivated much of the interest in the former is remarkably well realized in La$_2$NiO$_3$F. 
Specifically, we find that this material is essentially a single-band system with a sharply 2D Fermi surface dominated by the Ni-3$d_{x^2-y^2}$ states. Besides, compared to infinite-layer nickelates (where Ni$^{1+}$ is not Cu$^{2+}$ \cite{pickett-prb04}), the additional features of electronic structure much better match the cuprate-like picture, in particular the charge transfer energy which is 3.6~eV in La$_2$NiO$_3$F. 
This energy is considerably reduced compared to the infinite-layer superconducting nickelates \cite{botana20prx,cano20b,lechermann20}, and similar to the $n=3$ trilayer systems \cite{botana20trilayer}. 
We spot the fluorite spacer in La$_2$NiO$_3$F as an important ingredient to obtain these features. Its rather ionic character, in particular, increases the covalency within the NiO$_2$ layers and drastically enhaces their effective decoupling. 
Thus, this $n=1$ T'-type structure emerges as a very interesting setup where cuprate-like features are promoted with the Ni atom getting close to the ideal Ni$^{1+}$.

\section{Computational methods}

The electronic-structure calculations were performed using the all-electron code {\sc{WIEN2k}}~\cite{WIEN2k} based on the full-potential augmented plane-wave plus local orbitals method (APW+LO). 
We used the structural parameters determined experimentally \cite{clemens20,kobler91} and the Perdew-Burke-Ernzerhof (PBE) form of the generalized gradient approximation (GGA)~\cite{PBE} for the nonmagnetic calculations. 
We used muffin-tin radii of 2.5, 2.15, and 1.60 a.u. for the La (Nd), Ni, and O (F) atoms, respectively, and a plane-wave cutoff $R_{\rm MT}K_{\rm max}$ = 7.0. 
In the case of Nd, we treated the 4$f$-states as core electrons. We employed both conventional (2-formula-units) and primitive (1-formula-unit) units cells for the nonmagnetic calculations. 
The integration over the Brillouin zone was done using a Monkhorst-Pack mesh of $18 \times 18 \times 5$ ($12 \times 12 \times 12$) $k$-points for the self-consistent calculations, while a denser $24 \times 24 \times 8$ ($36 \times 36 \times 36 $) $k$-mesh was used to compute the Fermi surface within the conventional (primitive) cell. 
Further, the maximally localized Wannier functions (MLWFs) \cite{MLWF} were calculated by interfacing the WANNIER90 package \cite{Wannier90} to WIEN2k using the WIEN2WANNIER code \cite{Wien2wannier}.

The main calculations were performed considering the O-F arrangement within the fluorite spacers illustrated in Fig. \ref{f:0}(a). This is the high-symmetry configuration that minimizes the overall energy within the conventional unit cell. The configuration in which the O-F alternation is not rotated 90$^\circ$ along $c$, however, is only 5~meV/f.u higher in energy and produces essentially the same results.
We also considered the La$_2$OF fluorite spacer within the virtual crystal approximation (VCA) as in \cite{arita20prb}. This method, however, fails to reproduce the enhanced electronegativity of the F atom in our system. This circumstance manifests, in particular, in substantially different results for the $p$ states. We then concluded that, to the purpose of our analysis, VCA is not reliable and consequently we stuck to `real crystal' calculations.  

In addition, we performed magnetic calculations using both PBE and the local density approximation (LDA)~\cite{LDA} as the former is known to overestimate the tendency towards magnetic order, especially in metals. The magnetic solutions were obtained using different cells according to the magnetic order under consideration. For the ferromagnetic (FM) order we used the primitive body-centered tetragonal cell (BCT), while for the A-type
antiferromagnetic (AFM) order with the spins adjacent NiO$_2$ planes pointing in opposite directions ---{\it i.e.} FM planes stacked AFM along $c$--- we used the conventional tetragonal cell encompassing 2 formula units. 
For the C-type AFM order ---{\it i.e.} checkerboard order in-plane --- 
we used a monoclinic cell encompassing 2 formula units.
In addition, we also investigated the FM vs AFM vertical stacking of such a C-type in-plane order using a 4-formula-unit tetragonal cell.
Finally, we also considered the E-type AFM order where spins are aligned forming a double stripe structure using a body-centered orthorhombic cell encompassing 4 formula units. 

\section{Results}

\subsection{Nonmagnetic electronic structure}

\begin{figure}[b!]
    \includegraphics[width=.475\textwidth]{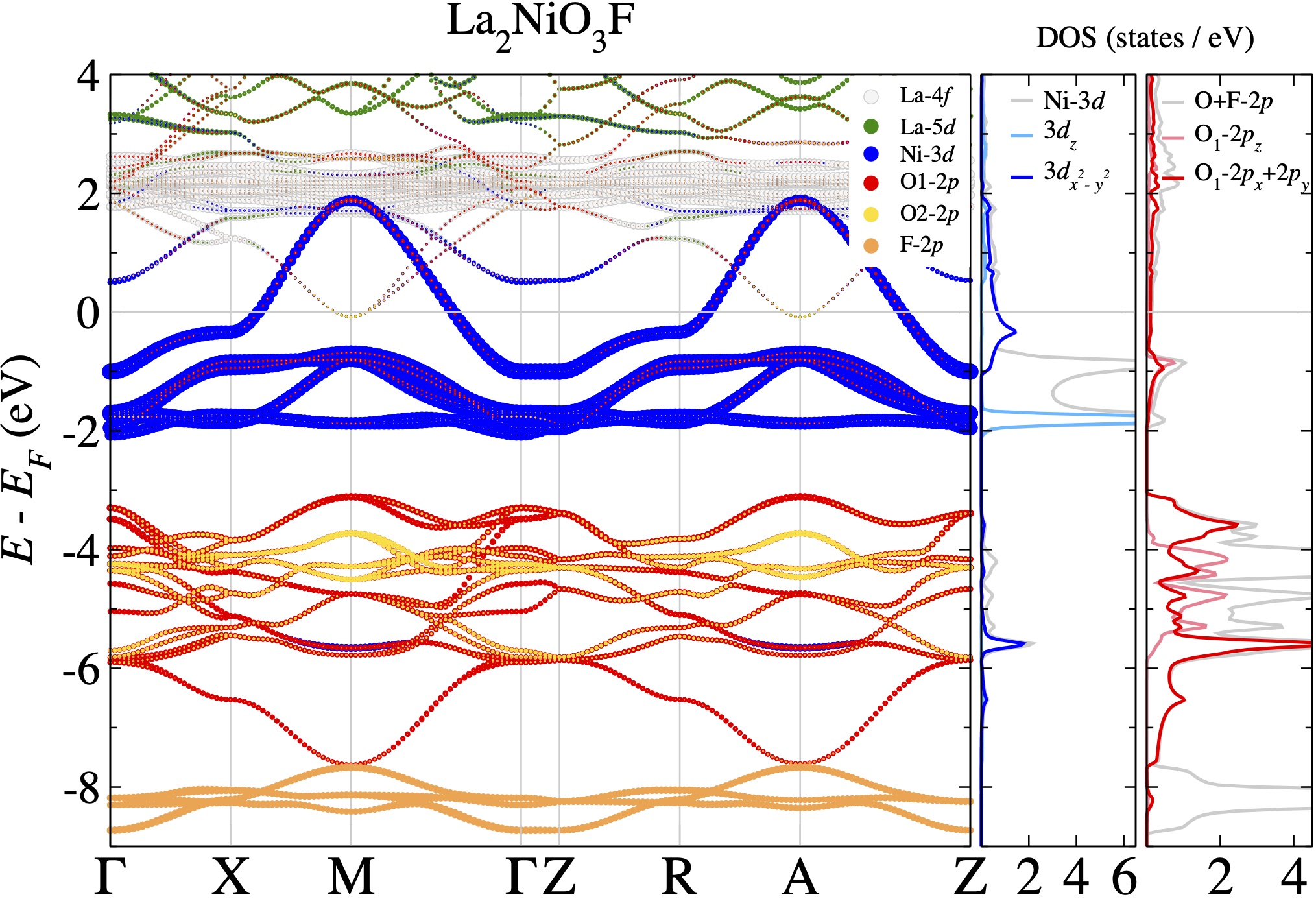}
    \caption{
    Band structure and orbital resolved density of states (DOS) of La$_2$NiO$_3$F ($I4/mmm$ space group $a = 3.9925$~\AA, $c = 12.5150$~\AA \cite{clemens20}). O1 refers to the oxygens in the NiO$_2$ layer, while O2 to the fluorite block.}
    \label{f:1}
\end{figure}

Fig. \ref{f:1} shows the band structure and orbital-resolved density of states (DOS) of La$_2$NiO$_3$F. 
The main features near the Fermi level are associated to the Ni-3$d$ bands. In fact, there is a clear splitting between the Ni-3$d$ and the O/F-2$p$ states with a gap of $\sim 1$~eV between these two manifolds. The O-2$p$ bands, in particular, extend below the Fermi energy from $-3$~eV to $-7.5$~eV, while the F-2$p$ ones are further below at $\sim -8$~eV. Yet, the DOS reveals some degree of hybridization between the Ni-3$d_{x^2-y^2}$ and the O-2$p$ states.
The bands above the Fermi energy derive from the La states mainly. One of them dips down and just touch the Fermi level at M and A. 
This provides an incipient self-doping of the Ni-3$d$ Fermi surface that, however, is different compared to the infinite-layer case \cite{pickett-prb04}. In La$_2$NiO$_3$F, in particular, no La-5$d_{z^2}$ states at $\Gamma$ are involved. 
The different self-doping is due the suppressed hybridization along $c$ resulting from the presence of the fluorite spacer between the NiO$_2$ planes. This situation is similar to that of the trilayer (hetero-)structures \cite{cano20c,botana20trilayer}. In the present case, however, the enhanced electronegativity of the F atom ---and thereby the enhanced ionicity of the fluorite spacer--- represents an additional key ingredient that suppresses such a hybridization and weakens the overall self-doping effect.   

\begin{figure}[t!]
    \includegraphics[width=.49\textwidth]{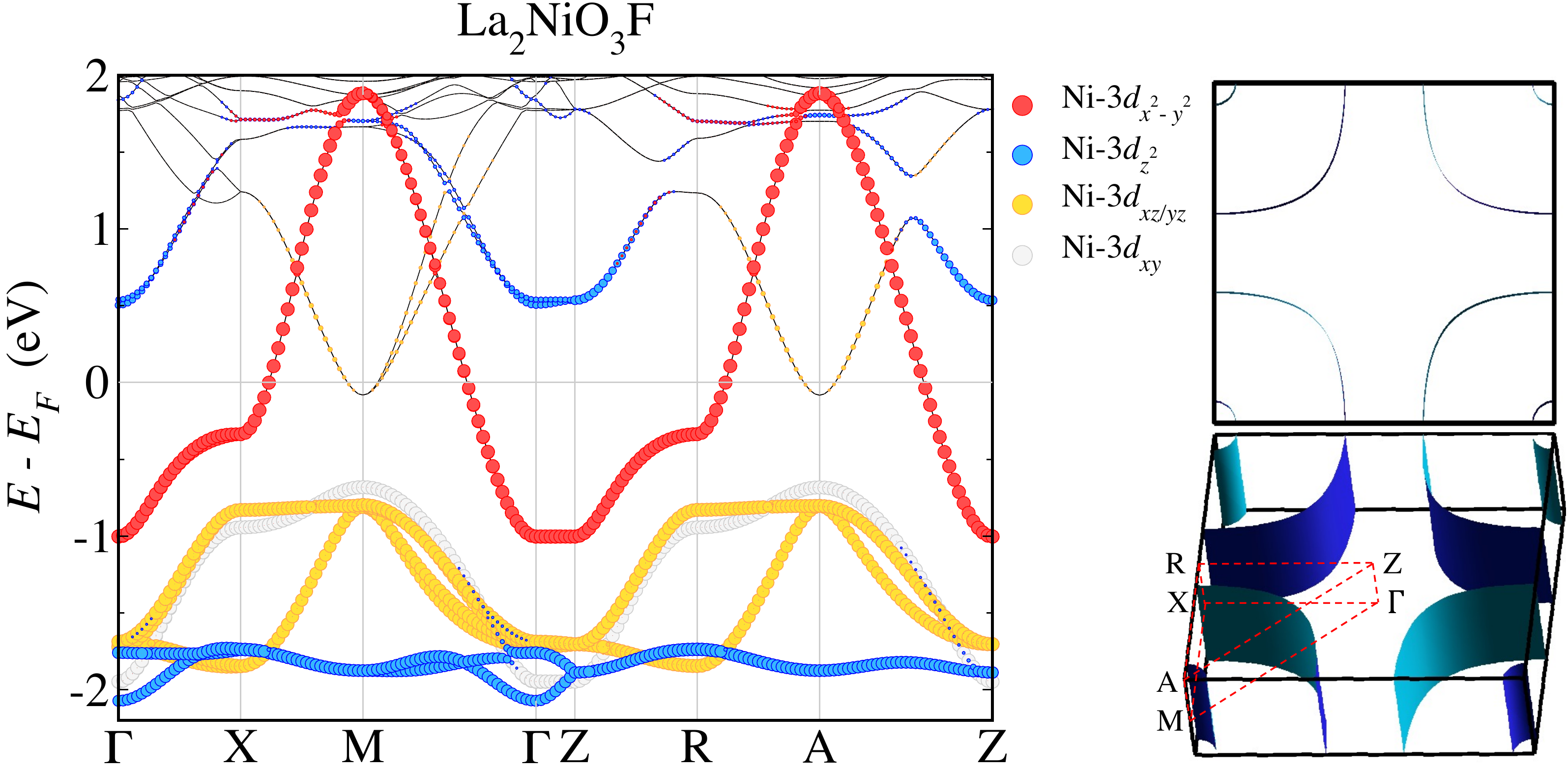}
    \caption{Band structure near the Fermi energy and Fermi surface of La$_2$NiO$_3$F (top and perspective views).}
    \label{f:2}
\end{figure}

Fig. \ref{f:2} highlights the different Ni-3$d$ contributions to the bands near the Fermi level and the corresponding Fermi surface. The band crossing the Fermi level is associated to the $d_{x^2-y^2}$ states. The $d_{z^2}$ band, in its turn, is located 2~eV below the Fermi energy and is remarkably flat all along the Brillouin zone, giving rise to a sharp feature in the DOS [see Fig. \ref{f:1}].  
The $d_{xz/yz}$ and $d_{xy}$ bands then appear sandwiched between the $e_{g}=\{d_{x^2-y^2},d_{z^2}\}$ states. 
The Fermi surface, in its turn, is remarkably two-dimensional (2D) and wraps the M-A line as can be seen in Fig. \ref{f:2}, including the small electron pocket.   
This exceptionally 2D character as well as the flatness of the $d_{z^2}$ is again due to the suppressed hybridization along $c$ (due to the fluorite spacer containing the highly electronegative F atom).

\subsection{Tight binding \& Wannierization}

The main Ni-3$d_{x^2-y^2}$ band can be reproduced using a simple one-band tight-binding model. The parameters of the model are summarized in Table \ref{t:tb}.
Compared to the infinite-layer nickelates \cite{pickett-prb04,botana20prx}, the out-of-plane hopping plays a minor role due to the exceptional 2D character of this $d_{x^2-y^2}$ band in La$_2$NiO$_3$F.
Otherwise, the values of the in-plane hopping integrals are similar. In particular, the ratio between longer-range to nearest-neighbor hopping defined as $t'/t = (|t_2| + |t_3|)/|t_1|$ turns out to be 0.41 (vs 0.37 in the infinite-layer case \cite{botana20prx}).
  
\begin{table}[b!]
\begin{tabular}{r l p{1ex} l}
\hline \hline 
$t_n$ & (meV) && $f_n({\mathbf k})$ \\
\hline
213&&& 1 \\
$-352$ &&& $2 [\cos(k_x a) +  \cos(k_y a)]$ \\
98 &&& $4 \cos(k_x a)   \cos(k_y a)$ \\
$-45$ &&& $2 [\cos(2 k_x a) +  \cos(2 k_y a)]$\\
$-116$ &&& $v^2({\mathbf k}) \cos({k_x a\over 2})\cos({k_y b\over 2})\cos({k_z c\over 2})$ \\
\hline \hline
\end{tabular}
\caption{Tight-binding fit of the main Ni-3$d_{x^2-y^2}$ band near the Fermi level with $\epsilon({\mathbf k}) = \sum_n t_n f_n ({\mathbf k})$ and $v({\mathbf k}) = [\cos(k_x a) -  \cos(k_y a)]/2$.}
\label{t:tb}
\end{table}

To gain further insight on the electronic properties of La$_2$NiO$_3$F we performed an analysis based on MLWFs.
The Ni-3$d$ manifold of the band structure can be reasonably reproduced by means of MLWFs obtained out of five Ni-centered $d$ orbitals. However, to reproduce the extra features near the Fermi level ({\it i.e.} the self-doping effect), not only La-$d$ orbitals but also O1-$p$ ones need to be included. 
Inclusion of O2-$p$, F-$p$ and La-$f$ orbitals improves the overall fit. Here and hereafter O1 refers to the oxygens in the NiO$_2$ layer while O2 to the fluorite spacer.

The on-site energies and hoppings obtained from the Wannier fits are summarized in Table \ref{tab:wann}. The splitting between the $d_{x^2 -y^2}$ and $d_{z^2}$ energies is $\delta \epsilon_{e_g}=0.49$~eV. This splitting is in between the one obtained for the infinite- and tri-layer systems, all of them smaller than the typical splitting of the cuprates. 
We note, however, that the distribution of $d_{z^2}$ states is bipartite in the present case as there is a gap between fully occupied and fully empty $d_{z^2}$ states (see Figs. \ref{f:1} and \ref{f:2}). On the other hand, the charge transfer energy $\Delta = \epsilon_d - \epsilon_p$ is 3.6~eV ($\Delta$ refers to $d_{x^2-y^2}$ and $p_x$). This energy is largely reduced compared to the infinite-layer systems, and turns out to be essentially the same charge transfer energy of the trilayer materials. The hopping parameters, in their turn, are almost identical for the $d_{x^2-y^2}$-$p_x$ case while some differences are observed in the other hoppings. 

\begin{table}[t!]
 \begin{tabular}{p{7em} r p{4em} p{6em} r}
\hline \hline
\multicolumn{5}{l}{Wannier on-site energies (eV)} \\
\hline
$d_{x^2-y^2}$ &    $-$1.06   && &  \\ 
$d_{z^2}$     &    $-$1.55   && $p_{x}$ (O1)   &     $-$4.65  \\  
$d_{xz/yz}$   &    $-$1.49   && $p_{y}$ (O1)   &     $-$3.28  \\
$d_{xy}$      &    $-$1.57   && $p_{z}$ (O1)   &     $-$3.30  \\
\hline
$p_{x}$ (O2)  &  $-$3.08   && $p_{x}$ (F) & $-$7.17  \\  
$p_{y}$ (O2)  &  $-$3.08   && $p_{y}$ (F) & $-$7.17  \\
$p_{z}$ (O2)  &  $-$3.01   && $p_{z}$ (F) & $-$7.20  \\
\hline 
\multicolumn{5}{l}{Wannier hoppings (eV)}   \\
\hline
$d_{x^2-y^2}$\,-\,$p_{x}$ (O1)  &  $-$1.22 &&&\\
$d_{z^2}$\,-\,$p_{x}$ (O1)  &  $-$0.44 &&$d_{z^2}$\,-\,$d_{z^2}$ (La)  &  $-$0.11 \\
$d_{xz/yz}$\,-\,$p_{z}$ (O1)  &   0.48 &&& \\
$d_{xy}$\,-\,$p_{y}$ (O1)  &   0.65 &&& \\
\hline \hline
\end{tabular}
\caption{Calculated on-site energies and hoppings for La$_2$NiO$_3$F derived from 
the Wannier functions. O1 is in the NiO$_2$ layer while O2 is the fluorite spacer. 
}
\label{tab:wann}
\end{table}

\subsection{Comparison with other nickelates and cuprates}

The low-energy features of La$_2$NiO$_3$F described above make this system a rather ideal $d^9$ material. This system is essentially a single $d_{x^2-y^2}$-band system like the cuprates, with a remarkable 2D character. 

In addition, the splitting between the $d_{x^2-y^2}$ and $d_{z^2}$ bands increases compared with the infinite-layer nickelates. 
The $e_g$ splitting has been discussed in relation to the superconducting $T_c$ in cuprates, where a larger splitting correlates to a higher $T_c$ due to the reduced mixing of these states. In La$_2$NiO$_3$F the $e_g$ splitting is neat below the Fermi level. However, there is another $d_{z^2}$ band above the Fermi level that crosses the $d_{x^2-y^2}$ one. We note that this situation is similar in the trilayer nickelates. 
In fact, the overall occupation of the Ni-3$d_{z^2}$ states is 0.76 per spin. However, the contribution of these states to the DOS at the Fermi level is negligible so that no effective $d_{x^2-y^2}$-$d_{z^2}$ mixing occurs for low-energy physics.

The occupation of the Ni-3$d_{x^2-y^2}$ orbitals, in its turn, is 0.6 per spin. 
Thus, the system can in principle be driven towards half-filling by means of hole doping (e.g. La~$\to$~Sr or O~$\to$~F substitutions). Hole doping can also be expected to wipe out the incipient self-doping of the main $d_{x^2-y^2}$ Fermi surface. Thus, an ideal cuprate-like situation may be better realized in La$_2$NiO$_3$F than in the infinite-layer nickelates. Besides, the ratio $t'/t$ describing the relative strength of longer-range to nearest-neighbor hopping is 0.41. This ratio, which is considered as an indicator of superconductivity \cite{andersen01-prl,aoki12-prb}, thus takes a value that is comparable to that of the cuprates with the highest $T_c$.  

Beyond that, the key parameters resulting from the Wannier fit are almost identical to those of the trilayer nickelates ($n=3$). In particular, the charge transfer energy is $\Delta = 3.6$~eV, and hence considerably lower than in the infinite-layer case. At the same time, the hybridization between the Ni-3$d_{x^2-y^2}$ and the O-2$p$ states is slightly lower yet similar to the trilayer case. Note that using a $t$-$J$ strong-coupling model based on these trilayer parameters, the superconducting $T_c$ has been predicted to reach 90~K \cite{botana20trilayer}.

\begin{figure}[!t]
    \includegraphics[width=.45\textwidth]{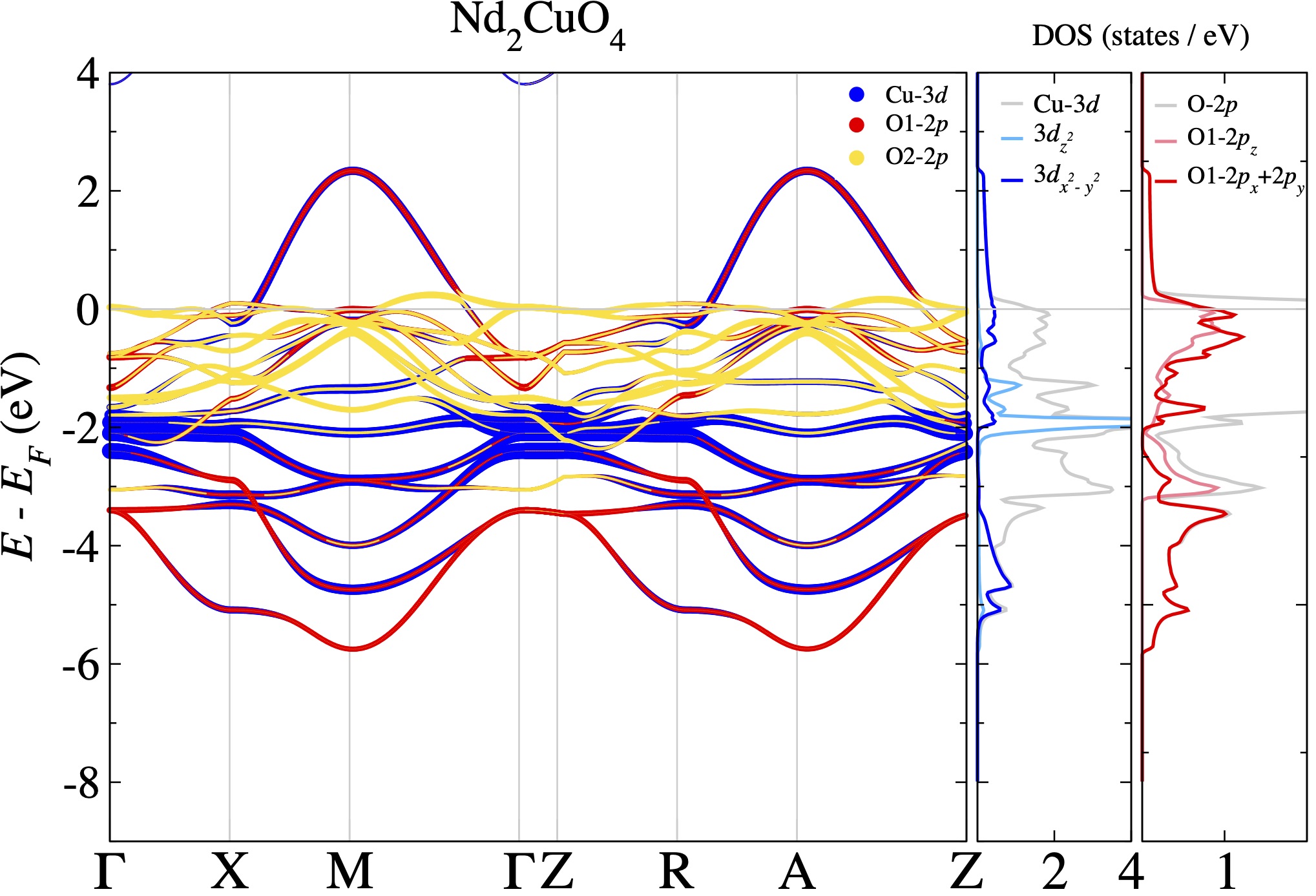}
    \caption{Band structure and orbital resolved density of states (DOS) of  Nd$_2$CuO$_4$. O1 refers to the oxygens in the CuO$_2$ layer while O2 to the fluorite block.}
    \label{f:Nd2CuO4}
\end{figure} 

It is worth noting that the interesting properties of La$_2$NiO$_3$F, while genuinely related to the ideal Ni$^{1+}$, do not match those of isostructural T'-type cuprates. Nd$_2$CuO$_4$ represents the natural system to compare with. This compound is reference material for the electron-doped cuprates, which are substantially less correlated systems than previously assumed \cite{jang16sr}. 
The electronic structure of Nd$_2$CuO$_4$ at the DFT level is shown in Fig. \ref{f:Nd2CuO4}. 
In contrast to the nickelates and other cuprates, this $n=1$ T'-type system features a direct overlap between $p$ and $d$ states \cite{botana20prx}. 
This overlap includes the O2-$p$ states from the fluorite block, which then cannot be considered as a mere spacer in this system. We analyzed the hypothetical series Nd$_2$CuO$_4$~$\to $~BaNdCuO$_3$F~$\to \dots$~Cs$_2$CuF$_4$ in the same T'-type ideal structure and found that this feature is robust, even if the fluorite block also contains F as in La$_2$NiO$_3$F. 
This reveals a rather fundamental difference between nickelates and cuprates in terms of $p$-$d$ hybridization (or, equivalently, charge transfer energies). Beyond that, this comparison shows that, contrary to what could have been extrapolated, $n=1$ T'-type nickelates do provide an additional platform to promote the electronic properties originally sought out in Ni$^{1+}$-hosting materials.

\subsection{Magnetism}

The magnetic ground state corresponds to the C-AFM order, in both LDA and PBE, among the different solutions that we considered. The energy difference with respect to the non-magnetic solution is 13~meV/Ni (80~meV/Ni) with LDA (PBE) exchange and correlation, while magnetic moment of the Ni atom is 0.57~$\mu_{B}$ (0.70~$\mu_{B}$). This energy gain is similarly small to that in the infinite-layer case and the resulting state is likewise metallic 
---as opposed to the AFM insulator obtained for CaCuO$_2$ \cite{botana20prx}. 
In addition, we found that the FM and AFM stackings of the C-AFM solution along $c$ in La$_2$NiO$_3$F are essentially degenerate in energy due to the enhanced 2D character of this system, which may lead to magnetic disorder.    
At the same time, we did not find neither FM nor A-AFM solutions for La$_2$NiO$_3$F, while these solutions exist for in the infinite-layer case. This means that the ferromagnetic order within the NiO$_2$ planes is strongly unfavorable. Instead, we found a small tendency towards E-AFM and SS-AFM with an energy difference with respect to the nonmagnetic state of 0.4~meV/Ni in LDA for the former (the latter is absent in LDA), and 39~meV/Ni and 9~meV/Ni respectively in PBE. 
Experimentally, some tendency towards magnetic order has been discussed although no robust conclusion can be drawn from the available data \cite{clemens20}. The theory vs experiments situation is therefore similar to that of the infinite-layer case and hence calls for additional details.

\section{Conclusions}

We have analyzed the electronic properties of La$_2$NiO$_3$F. This system provides the first realization of a $n=1$ T'-type nickelate with the same $d^9$ formal configuration of the superconducting infinite-layer case. We have found that the single-layer La$_2$NiO$_3$F is essentially a single-band system that much better materializes the intended analogy to cuprates, as originally sought from Ni$^{1+}$. 
In La$_2$NiO$_3$F, the NiO$_2$ layers are remarkably decoupled due to the F-containing fluorite spacer. This gives rise to a true 2D Fermi surface dominated by the Ni-3$d_{x^2-y^2}$ states. At the same time, the $d_{x^2-y^2}$-$d_{z^2}$ mixing is practically absent near the Fermi level.   
Besides, the ratio between longer-range to nearest-neighbor hopping slightly increases while the charge transfer energy decreases compared to the infinite-layer nickelates. The latter is similar to the trilayer case, where a superconducting $T_c$ reaching 90~K has recently been predicted.

Our results thus indicate  La$_2$NiO$_3$F as very promising candidate for superconductivity according to the above set of electronic properties. 
These properties may be further optimized in other members of this new single-layer family and/or tuned by means of control parameters such as applied pressure or epitaxial strain. These systems can thus be used to further examine the above quantities as descriptors of high-$T_c$ superconductivity, and thereby to clarify the underlying mechanisms. In fact, the global behavior of these systems ---beyond the potential emergence of superconductivity in unprecedented nickelates--- may eventually reveal different degrees of electronic correlations, which is however hard to establish from a pure {\it ab-initio} basis.
This broadened panorama is therefore expected to motivate further investigations on this class of quantum materials.

\vspace{1em}
\noindent {\it Acknowledgments.---} We thank V. Olevano and X. Blase for useful discussions and ANR-18-CE30-0018 for support.

\bibliography{bib.bib}

\begin{thebibliography}{38}%
\makeatletter
\providecommand \@ifxundefined [1]{%
 \@ifx{#1\undefined}
}%
\providecommand \@ifnum [1]{%
 \ifnum #1\expandafter \@firstoftwo
 \else \expandafter \@secondoftwo
 \fi
}%
\providecommand \@ifx [1]{%
 \ifx #1\expandafter \@firstoftwo
 \else \expandafter \@secondoftwo
 \fi
}%
\providecommand \natexlab [1]{#1}%
\providecommand \enquote  [1]{``#1''}%
\providecommand \bibnamefont  [1]{#1}%
\providecommand \bibfnamefont [1]{#1}%
\providecommand \citenamefont [1]{#1}%
\providecommand \href@noop [0]{\@secondoftwo}%
\providecommand \href [0]{\begingroup \@sanitize@url \@href}%
\providecommand \@href[1]{\@@startlink{#1}\@@href}%
\providecommand \@@href[1]{\endgroup#1\@@endlink}%
\providecommand \@sanitize@url [0]{\catcode `\\12\catcode `\$12\catcode
  `\&12\catcode `\#12\catcode `\^12\catcode `\_12\catcode `\%12\relax}%
\providecommand \@@startlink[1]{}%
\providecommand \@@endlink[0]{}%
\providecommand \url  [0]{\begingroup\@sanitize@url \@url }%
\providecommand \@url [1]{\endgroup\@href {#1}{\urlprefix }}%
\providecommand \urlprefix  [0]{URL }%
\providecommand \Eprint [0]{\href }%
\providecommand \doibase [0]{http://dx.doi.org/}%
\providecommand \selectlanguage [0]{\@gobble}%
\providecommand \bibinfo  [0]{\@secondoftwo}%
\providecommand \bibfield  [0]{\@secondoftwo}%
\providecommand \translation [1]{[#1]}%
\providecommand \BibitemOpen [0]{}%
\providecommand \bibitemStop [0]{}%
\providecommand \bibitemNoStop [0]{.\EOS\space}%
\providecommand \EOS [0]{\spacefactor3000\relax}%
\providecommand \BibitemShut  [1]{\csname bibitem#1\endcsname}%
\let\auto@bib@innerbib\@empty
\bibitem [{\citenamefont {Botana}\ \emph {et~al.}(2021)\citenamefont {Botana},
  \citenamefont {Bernardini},\ and\ \citenamefont
  {Cano}}]{review-nickelates20}%
  \BibitemOpen
  \bibfield  {author} {\bibinfo {author} {\bibfnamefont {A.~S.}\ \bibnamefont
  {Botana}}, \bibinfo {author} {\bibfnamefont {F.}~\bibnamefont {Bernardini}},
  \ and\ \bibinfo {author} {\bibfnamefont {A.}~\bibnamefont {Cano}},\
  }\href@noop {} {\bibfield  {journal} {\bibinfo  {journal} {JETP}\ }\textbf
  {\bibinfo {volume} {159}},\ \bibinfo {pages} {711} (\bibinfo {year}
  {2021})},\ \Eprint {http://arxiv.org/abs/2012.02764} {arXiv:2012.02764}
  \BibitemShut {NoStop}%
\bibitem [{\citenamefont {Norman}(2020)}]{norman20-p}%
  \BibitemOpen
  \bibfield  {author} {\bibinfo {author} {\bibfnamefont {M.~R.}\ \bibnamefont
  {Norman}},\ }\href@noop {} {\bibfield  {journal} {\bibinfo  {journal}
  {Physics}\ }\textbf {\bibinfo {volume} {13}},\ \bibinfo {pages} {85}
  (\bibinfo {year} {2020})}\BibitemShut {NoStop}%
\bibitem [{\citenamefont {Pickett}(2021)}]{pickett20-np}%
  \BibitemOpen
  \bibfield  {author} {\bibinfo {author} {\bibfnamefont {W.~E.}\ \bibnamefont
  {Pickett}},\ }\href@noop {} {\bibfield  {journal} {\bibinfo  {journal}
  {Nature Reviews Physics}\ }\textbf {\bibinfo {volume} {3}},\ \bibinfo {pages}
  {7} (\bibinfo {year} {2021})}\BibitemShut {NoStop}%
\bibitem [{\citenamefont {Crespin}\ \emph {et~al.}(1983)\citenamefont
  {Crespin}, \citenamefont {Levitz},\ and\ \citenamefont
  {Gatineau}}]{crespin83}%
  \BibitemOpen
  \bibfield  {author} {\bibinfo {author} {\bibfnamefont {M.}~\bibnamefont
  {Crespin}}, \bibinfo {author} {\bibfnamefont {P.}~\bibnamefont {Levitz}}, \
  and\ \bibinfo {author} {\bibfnamefont {L.}~\bibnamefont {Gatineau}},\ }\href
  {\doibase 10.1039/F29837901181} {\bibfield  {journal} {\bibinfo  {journal}
  {J. Chem. Soc.{,} Faraday Trans. 2}\ }\textbf {\bibinfo {volume} {79}},\
  \bibinfo {pages} {1181} (\bibinfo {year} {1983})}\BibitemShut {NoStop}%
\bibitem [{\citenamefont {Hayward}\ \emph {et~al.}(1999)\citenamefont
  {Hayward}, \citenamefont {Green}, \citenamefont {Rosseinsky},\ and\
  \citenamefont {Sloan}}]{hayward99}%
  \BibitemOpen
  \bibfield  {author} {\bibinfo {author} {\bibfnamefont {M.}~\bibnamefont
  {Hayward}}, \bibinfo {author} {\bibfnamefont {M.}~\bibnamefont {Green}},
  \bibinfo {author} {\bibfnamefont {M.}~\bibnamefont {Rosseinsky}}, \ and\
  \bibinfo {author} {\bibfnamefont {J.}~\bibnamefont {Sloan}},\ }\href@noop {}
  {\bibfield  {journal} {\bibinfo  {journal} {Journal of the American Chemical
  Society}\ }\textbf {\bibinfo {volume} {121}},\ \bibinfo {pages} {8843}
  (\bibinfo {year} {1999})}\BibitemShut {NoStop}%
\bibitem [{\citenamefont {Hayward}\ and\ \citenamefont
  {Rosseinsky}(2003)}]{hayward2003}%
  \BibitemOpen
  \bibfield  {author} {\bibinfo {author} {\bibfnamefont {M.}~\bibnamefont
  {Hayward}}\ and\ \bibinfo {author} {\bibfnamefont {M.}~\bibnamefont
  {Rosseinsky}},\ }\href {\doibase
  https://doi.org/10.1016/S1293-2558(03)00111-0} {\bibfield  {journal}
  {\bibinfo  {journal} {Solid State Sciences}\ }\textbf {\bibinfo {volume}
  {5}},\ \bibinfo {pages} {839 } (\bibinfo {year} {2003})}\BibitemShut
  {NoStop}%
\bibitem [{\citenamefont {{Crespin}}\ \emph {et~al.}(2005)\citenamefont
  {{Crespin}}, \citenamefont {{Isnard}}, \citenamefont {{Dubois}},
  \citenamefont {{Choisnet}},\ and\ \citenamefont {{Odier}}}]{crespin05}%
  \BibitemOpen
  \bibfield  {author} {\bibinfo {author} {\bibfnamefont {M.}~\bibnamefont
  {{Crespin}}}, \bibinfo {author} {\bibfnamefont {O.}~\bibnamefont {{Isnard}}},
  \bibinfo {author} {\bibfnamefont {F.}~\bibnamefont {{Dubois}}}, \bibinfo
  {author} {\bibfnamefont {J.}~\bibnamefont {{Choisnet}}}, \ and\ \bibinfo
  {author} {\bibfnamefont {P.}~\bibnamefont {{Odier}}},\ }\href {\doibase
  10.1016/j.jssc.2005.01.023} {\bibfield  {journal} {\bibinfo  {journal}
  {Journal of Solid State Chemistry France}\ }\textbf {\bibinfo {volume}
  {178}},\ \bibinfo {pages} {1326} (\bibinfo {year} {2005})}\BibitemShut
  {NoStop}%
\bibitem [{\citenamefont {Poltavets}\ \emph {et~al.}(2007)\citenamefont
  {Poltavets}, \citenamefont {Lokshin}, \citenamefont {Croft}, \citenamefont
  {Mandal}, \citenamefont {Egami},\ and\ \citenamefont
  {Greenblatt}}]{poltavets1}%
  \BibitemOpen
  \bibfield  {author} {\bibinfo {author} {\bibfnamefont {V.~V.}\ \bibnamefont
  {Poltavets}}, \bibinfo {author} {\bibfnamefont {K.~A.}\ \bibnamefont
  {Lokshin}}, \bibinfo {author} {\bibfnamefont {M.}~\bibnamefont {Croft}},
  \bibinfo {author} {\bibfnamefont {T.~K.}\ \bibnamefont {Mandal}}, \bibinfo
  {author} {\bibfnamefont {T.}~\bibnamefont {Egami}}, \ and\ \bibinfo {author}
  {\bibfnamefont {M.}~\bibnamefont {Greenblatt}},\ }\href {\doibase
  10.1021/ic701480v} {\bibfield  {journal} {\bibinfo  {journal} {Inorg. Chem.}\
  }\textbf {\bibinfo {volume} {46}},\ \bibinfo {pages} {10887} (\bibinfo {year}
  {2007})}\BibitemShut {NoStop}%
\bibitem [{\citenamefont {Poltavets}\ \emph {et~al.}(2006)\citenamefont
  {Poltavets}, \citenamefont {Lokshin}, \citenamefont {Dikmen}, \citenamefont
  {Croft}, \citenamefont {Egami},\ and\ \citenamefont
  {Greenblatt}}]{poltavets2}%
  \BibitemOpen
  \bibfield  {author} {\bibinfo {author} {\bibfnamefont {V.~V.}\ \bibnamefont
  {Poltavets}}, \bibinfo {author} {\bibfnamefont {K.~A.}\ \bibnamefont
  {Lokshin}}, \bibinfo {author} {\bibfnamefont {S.}~\bibnamefont {Dikmen}},
  \bibinfo {author} {\bibfnamefont {M.}~\bibnamefont {Croft}}, \bibinfo
  {author} {\bibfnamefont {T.}~\bibnamefont {Egami}}, \ and\ \bibinfo {author}
  {\bibfnamefont {M.}~\bibnamefont {Greenblatt}},\ }\href
  {https://doi.org/10.1021/ja063031o} {\bibfield  {journal} {\bibinfo
  {journal} {Journal of the American Chemical Society}\ }\textbf {\bibinfo
  {volume} {128}},\ \bibinfo {pages} {9050} (\bibinfo {year}
  {2006})}\BibitemShut {NoStop}%
\bibitem [{\citenamefont {Poltavets}\ \emph {et~al.}(2009)\citenamefont
  {Poltavets}, \citenamefont {Greenblatt}, \citenamefont {Fecher},\ and\
  \citenamefont {Felser}}]{Poltavets_3}%
  \BibitemOpen
  \bibfield  {author} {\bibinfo {author} {\bibfnamefont {V.~V.}\ \bibnamefont
  {Poltavets}}, \bibinfo {author} {\bibfnamefont {M.}~\bibnamefont
  {Greenblatt}}, \bibinfo {author} {\bibfnamefont {G.~H.}\ \bibnamefont
  {Fecher}}, \ and\ \bibinfo {author} {\bibfnamefont {C.}~\bibnamefont
  {Felser}},\ }\href {\doibase 10.1103/PhysRevLett.102.046405} {\bibfield
  {journal} {\bibinfo  {journal} {Phys. Rev. Lett.}\ }\textbf {\bibinfo
  {volume} {102}},\ \bibinfo {pages} {046405} (\bibinfo {year}
  {2009})}\BibitemShut {NoStop}%
\bibitem [{\citenamefont {Poltavets}\ \emph {et~al.}(2010)\citenamefont
  {Poltavets}, \citenamefont {Lokshin}, \citenamefont {Nevidomskyy},
  \citenamefont {Croft}, \citenamefont {Tyson}, \citenamefont {Hadermann},
  \citenamefont {Van~Tendeloo}, \citenamefont {Egami}, \citenamefont {Kotliar},
  \citenamefont {ApRoberts-Warren}, \citenamefont {Dioguardi}, \citenamefont
  {Curro},\ and\ \citenamefont {Greenblatt}}]{Poltavets_4}%
  \BibitemOpen
  \bibfield  {author} {\bibinfo {author} {\bibfnamefont {V.~V.}\ \bibnamefont
  {Poltavets}}, \bibinfo {author} {\bibfnamefont {K.~A.}\ \bibnamefont
  {Lokshin}}, \bibinfo {author} {\bibfnamefont {A.~H.}\ \bibnamefont
  {Nevidomskyy}}, \bibinfo {author} {\bibfnamefont {M.}~\bibnamefont {Croft}},
  \bibinfo {author} {\bibfnamefont {T.~A.}\ \bibnamefont {Tyson}}, \bibinfo
  {author} {\bibfnamefont {J.}~\bibnamefont {Hadermann}}, \bibinfo {author}
  {\bibfnamefont {G.}~\bibnamefont {Van~Tendeloo}}, \bibinfo {author}
  {\bibfnamefont {T.}~\bibnamefont {Egami}}, \bibinfo {author} {\bibfnamefont
  {G.}~\bibnamefont {Kotliar}}, \bibinfo {author} {\bibfnamefont
  {N.}~\bibnamefont {ApRoberts-Warren}}, \bibinfo {author} {\bibfnamefont
  {A.~P.}\ \bibnamefont {Dioguardi}}, \bibinfo {author} {\bibfnamefont {N.~J.}\
  \bibnamefont {Curro}}, \ and\ \bibinfo {author} {\bibfnamefont
  {M.}~\bibnamefont {Greenblatt}},\ }\href {\doibase
  10.1103/PhysRevLett.104.206403} {\bibfield  {journal} {\bibinfo  {journal}
  {Phys. Rev. Lett.}\ }\textbf {\bibinfo {volume} {104}},\ \bibinfo {pages}
  {206403} (\bibinfo {year} {2010})}\BibitemShut {NoStop}%
\bibitem [{\citenamefont {Anisimov}\ \emph {et~al.}(1999)\citenamefont
  {Anisimov}, \citenamefont {Bukhvalov},\ and\ \citenamefont
  {Rice}}]{anisimov99}%
  \BibitemOpen
  \bibfield  {author} {\bibinfo {author} {\bibfnamefont {V.~I.}\ \bibnamefont
  {Anisimov}}, \bibinfo {author} {\bibfnamefont {D.}~\bibnamefont {Bukhvalov}},
  \ and\ \bibinfo {author} {\bibfnamefont {T.~M.}\ \bibnamefont {Rice}},\
  }\href {\doibase 10.1103/PhysRevB.59.7901} {\bibfield  {journal} {\bibinfo
  {journal} {Phys. Rev. B}\ }\textbf {\bibinfo {volume} {59}},\ \bibinfo
  {pages} {7901} (\bibinfo {year} {1999})}\BibitemShut {NoStop}%
\bibitem [{\citenamefont {Lee}\ and\ \citenamefont
  {Pickett}(2004)}]{pickett-prb04}%
  \BibitemOpen
  \bibfield  {author} {\bibinfo {author} {\bibfnamefont {K.-W.}\ \bibnamefont
  {Lee}}\ and\ \bibinfo {author} {\bibfnamefont {W.~E.}\ \bibnamefont
  {Pickett}},\ }\href {\doibase 10.1103/PhysRevB.70.165109} {\bibfield
  {journal} {\bibinfo  {journal} {Phys. Rev. B}\ }\textbf {\bibinfo {volume}
  {70}},\ \bibinfo {pages} {165109} (\bibinfo {year} {2004})}\BibitemShut
  {NoStop}%
\bibitem [{\citenamefont {Zhang}\ \emph {et~al.}(2017)\citenamefont {Zhang},
  \citenamefont {Botana}, \citenamefont {Freeland}, \citenamefont {Phelan},
  \citenamefont {Zheng}, \citenamefont {Pardo}, \citenamefont {Norman},\ and\
  \citenamefont {Mitchell}}]{zhang_nat_phys}%
  \BibitemOpen
  \bibfield  {author} {\bibinfo {author} {\bibfnamefont {J.}~\bibnamefont
  {Zhang}}, \bibinfo {author} {\bibfnamefont {A.~S.}\ \bibnamefont {Botana}},
  \bibinfo {author} {\bibfnamefont {J.~W.}\ \bibnamefont {Freeland}}, \bibinfo
  {author} {\bibfnamefont {D.}~\bibnamefont {Phelan}}, \bibinfo {author}
  {\bibfnamefont {H.}~\bibnamefont {Zheng}}, \bibinfo {author} {\bibfnamefont
  {V.}~\bibnamefont {Pardo}}, \bibinfo {author} {\bibfnamefont {M.~R.}\
  \bibnamefont {Norman}}, \ and\ \bibinfo {author} {\bibfnamefont {J.~F.}\
  \bibnamefont {Mitchell}},\ }\href {\doibase 10.1038/nphys4149} {\bibfield
  {journal} {\bibinfo  {journal} {Nature Physics}\ }\textbf {\bibinfo {volume}
  {13}},\ \bibinfo {pages} {864} (\bibinfo {year} {2017})}\BibitemShut
  {NoStop}%
\bibitem [{\citenamefont {{Li}}\ \emph {et~al.}(2019)\citenamefont {{Li}},
  \citenamefont {{Lee}}, \citenamefont {{Wang}}, \citenamefont {{Osada}},
  \citenamefont {{Crossley}}, \citenamefont {{Lee}}, \citenamefont {{Cui}},
  \citenamefont {{Hikita}},\ and\ \citenamefont {{Hwang}}}]{hwang19a}%
  \BibitemOpen
  \bibfield  {author} {\bibinfo {author} {\bibfnamefont {D.}~\bibnamefont
  {{Li}}}, \bibinfo {author} {\bibfnamefont {K.}~\bibnamefont {{Lee}}},
  \bibinfo {author} {\bibfnamefont {B.~Y.}\ \bibnamefont {{Wang}}}, \bibinfo
  {author} {\bibfnamefont {M.}~\bibnamefont {{Osada}}}, \bibinfo {author}
  {\bibfnamefont {S.}~\bibnamefont {{Crossley}}}, \bibinfo {author}
  {\bibfnamefont {H.~R.}\ \bibnamefont {{Lee}}}, \bibinfo {author}
  {\bibfnamefont {Y.}~\bibnamefont {{Cui}}}, \bibinfo {author} {\bibfnamefont
  {Y.}~\bibnamefont {{Hikita}}}, \ and\ \bibinfo {author} {\bibfnamefont
  {H.~Y.}\ \bibnamefont {{Hwang}}},\ }\href {\doibase
  10.1038/s41586-019-1496-5} {\bibfield  {journal} {\bibinfo  {journal} {\nat}\
  }\textbf {\bibinfo {volume} {572}},\ \bibinfo {pages} {624} (\bibinfo {year}
  {2019})}\BibitemShut {NoStop}%
\bibitem [{\citenamefont {Zeng}\ \emph {et~al.}(2020)\citenamefont {Zeng},
  \citenamefont {Tang}, \citenamefont {Yin}, \citenamefont {Li}, \citenamefont
  {Li}, \citenamefont {Huang}, \citenamefont {Hu}, \citenamefont {Liu},
  \citenamefont {Omar}, \citenamefont {Jani}, \citenamefont {Lim},
  \citenamefont {Han}, \citenamefont {Wan}, \citenamefont {Yang}, \citenamefont
  {Pennycook}, \citenamefont {Wee},\ and\ \citenamefont {Ariando}}]{ariando20}%
  \BibitemOpen
  \bibfield  {author} {\bibinfo {author} {\bibfnamefont {S.}~\bibnamefont
  {Zeng}}, \bibinfo {author} {\bibfnamefont {C.~S.}\ \bibnamefont {Tang}},
  \bibinfo {author} {\bibfnamefont {X.}~\bibnamefont {Yin}}, \bibinfo {author}
  {\bibfnamefont {C.}~\bibnamefont {Li}}, \bibinfo {author} {\bibfnamefont
  {M.}~\bibnamefont {Li}}, \bibinfo {author} {\bibfnamefont {Z.}~\bibnamefont
  {Huang}}, \bibinfo {author} {\bibfnamefont {J.}~\bibnamefont {Hu}}, \bibinfo
  {author} {\bibfnamefont {W.}~\bibnamefont {Liu}}, \bibinfo {author}
  {\bibfnamefont {G.~J.}\ \bibnamefont {Omar}}, \bibinfo {author}
  {\bibfnamefont {H.}~\bibnamefont {Jani}}, \bibinfo {author} {\bibfnamefont
  {Z.~S.}\ \bibnamefont {Lim}}, \bibinfo {author} {\bibfnamefont
  {K.}~\bibnamefont {Han}}, \bibinfo {author} {\bibfnamefont {D.}~\bibnamefont
  {Wan}}, \bibinfo {author} {\bibfnamefont {P.}~\bibnamefont {Yang}}, \bibinfo
  {author} {\bibfnamefont {S.~J.}\ \bibnamefont {Pennycook}}, \bibinfo {author}
  {\bibfnamefont {A.~T.~S.}\ \bibnamefont {Wee}}, \ and\ \bibinfo {author}
  {\bibfnamefont {A.}~\bibnamefont {Ariando}},\ }\href {\doibase
  10.1103/PhysRevLett.125.147003} {\bibfield  {journal} {\bibinfo  {journal}
  {Phys. Rev. Lett.}\ }\textbf {\bibinfo {volume} {125}},\ \bibinfo {pages}
  {147003} (\bibinfo {year} {2020})}\BibitemShut {NoStop}%
\bibitem [{\citenamefont {{Osada}}\ \emph {et~al.}(2020)\citenamefont
  {{Osada}}, \citenamefont {{Wang}}, \citenamefont {{Goodge}}, \citenamefont
  {{Lee}}, \citenamefont {{Yoon}}, \citenamefont {{Sakuma}}, \citenamefont
  {{Li}}, \citenamefont {{Miura}}, \citenamefont {{Kourkoutis}},\ and\
  \citenamefont {{Hwang}}}]{hwang20Pr-a}%
  \BibitemOpen
  \bibfield  {author} {\bibinfo {author} {\bibfnamefont {M.}~\bibnamefont
  {{Osada}}}, \bibinfo {author} {\bibfnamefont {B.~Y.}\ \bibnamefont {{Wang}}},
  \bibinfo {author} {\bibfnamefont {B.~H.}\ \bibnamefont {{Goodge}}}, \bibinfo
  {author} {\bibfnamefont {K.}~\bibnamefont {{Lee}}}, \bibinfo {author}
  {\bibfnamefont {H.}~\bibnamefont {{Yoon}}}, \bibinfo {author} {\bibfnamefont
  {K.}~\bibnamefont {{Sakuma}}}, \bibinfo {author} {\bibfnamefont
  {D.}~\bibnamefont {{Li}}}, \bibinfo {author} {\bibfnamefont {M.}~\bibnamefont
  {{Miura}}}, \bibinfo {author} {\bibfnamefont {L.~F.}\ \bibnamefont
  {{Kourkoutis}}}, \ and\ \bibinfo {author} {\bibfnamefont {H.~Y.}\
  \bibnamefont {{Hwang}}},\ }\href {\doibase 10.1021/acs.nanolett.0c01392}
  {\bibfield  {journal} {\bibinfo  {journal} {Nano Letters}\ }\textbf {\bibinfo
  {volume} {20}},\ \bibinfo {pages} {5735} (\bibinfo {year}
  {2020})}\BibitemShut {NoStop}%
\bibitem [{\citenamefont {{Osada}}\ \emph {et~al.}()\citenamefont {{Osada}},
  \citenamefont {{Wang}}, \citenamefont {{Lee}}, \citenamefont {{Li}},\ and\
  \citenamefont {{Hwang}}}]{hwang20Pr-b}%
  \BibitemOpen
  \bibfield  {author} {\bibinfo {author} {\bibfnamefont {M.}~\bibnamefont
  {{Osada}}}, \bibinfo {author} {\bibfnamefont {B.~Y.}\ \bibnamefont {{Wang}}},
  \bibinfo {author} {\bibfnamefont {K.}~\bibnamefont {{Lee}}}, \bibinfo
  {author} {\bibfnamefont {D.}~\bibnamefont {{Li}}}, \ and\ \bibinfo {author}
  {\bibfnamefont {H.~Y.}\ \bibnamefont {{Hwang}}},\ }\href@noop {} {\ }\Eprint
  {http://arxiv.org/abs/2010.16101} {arXiv:2010.16101} \BibitemShut {NoStop}%
\bibitem [{\citenamefont {{Gu}}\ \emph {et~al.}()\citenamefont {{Gu}},
  \citenamefont {{Li}}, \citenamefont {{Wan}}, \citenamefont {{Li}},
  \citenamefont {{Guo}}, \citenamefont {{Yang}}, \citenamefont {{Li}},
  \citenamefont {{Zhu}}, \citenamefont {{Pan}}, \citenamefont {{Nie}},\ and\
  \citenamefont {{Wen}}}]{wen20a}%
  \BibitemOpen
  \bibfield  {author} {\bibinfo {author} {\bibfnamefont {Q.}~\bibnamefont
  {{Gu}}}, \bibinfo {author} {\bibfnamefont {Y.}~\bibnamefont {{Li}}}, \bibinfo
  {author} {\bibfnamefont {S.}~\bibnamefont {{Wan}}}, \bibinfo {author}
  {\bibfnamefont {H.}~\bibnamefont {{Li}}}, \bibinfo {author} {\bibfnamefont
  {W.}~\bibnamefont {{Guo}}}, \bibinfo {author} {\bibfnamefont
  {H.}~\bibnamefont {{Yang}}}, \bibinfo {author} {\bibfnamefont
  {Q.}~\bibnamefont {{Li}}}, \bibinfo {author} {\bibfnamefont {X.}~\bibnamefont
  {{Zhu}}}, \bibinfo {author} {\bibfnamefont {X.}~\bibnamefont {{Pan}}},
  \bibinfo {author} {\bibfnamefont {Y.}~\bibnamefont {{Nie}}}, \ and\ \bibinfo
  {author} {\bibfnamefont {H.-H.}\ \bibnamefont {{Wen}}},\ }\href@noop {} {\
  }\Eprint {http://arxiv.org/abs/2006.13123} {arXiv:2006.13123} \BibitemShut
  {NoStop}%
\bibitem [{\citenamefont {Gawraczy{\'n}ski}\ \emph {et~al.}(2019)\citenamefont
  {Gawraczy{\'n}ski}, \citenamefont {Kurzyd{\l}owski}, \citenamefont {Ewings},
  \citenamefont {Bandaru}, \citenamefont {Gadomski}, \citenamefont {Mazej},
  \citenamefont {Ruani}, \citenamefont {Bergenti}, \citenamefont {Jaro{\'n}},
  \citenamefont {Ozarowski}, \citenamefont {Hill}, \citenamefont
  {Leszczy{\'n}ski}, \citenamefont {Tok{\'a}r}, \citenamefont {Derzsi},
  \citenamefont {Barone}, \citenamefont {Wohlfeld}, \citenamefont {Lorenzana},\
  and\ \citenamefont {Grochala}}]{lorenzana19}%
  \BibitemOpen
  \bibfield  {author} {\bibinfo {author} {\bibfnamefont {J.}~\bibnamefont
  {Gawraczy{\'n}ski}}, \bibinfo {author} {\bibfnamefont {D.}~\bibnamefont
  {Kurzyd{\l}owski}}, \bibinfo {author} {\bibfnamefont {R.~A.}\ \bibnamefont
  {Ewings}}, \bibinfo {author} {\bibfnamefont {S.}~\bibnamefont {Bandaru}},
  \bibinfo {author} {\bibfnamefont {W.}~\bibnamefont {Gadomski}}, \bibinfo
  {author} {\bibfnamefont {Z.}~\bibnamefont {Mazej}}, \bibinfo {author}
  {\bibfnamefont {G.}~\bibnamefont {Ruani}}, \bibinfo {author} {\bibfnamefont
  {I.}~\bibnamefont {Bergenti}}, \bibinfo {author} {\bibfnamefont
  {T.}~\bibnamefont {Jaro{\'n}}}, \bibinfo {author} {\bibfnamefont
  {A.}~\bibnamefont {Ozarowski}}, \bibinfo {author} {\bibfnamefont
  {S.}~\bibnamefont {Hill}}, \bibinfo {author} {\bibfnamefont {P.~J.}\
  \bibnamefont {Leszczy{\'n}ski}}, \bibinfo {author} {\bibfnamefont
  {K.}~\bibnamefont {Tok{\'a}r}}, \bibinfo {author} {\bibfnamefont
  {M.}~\bibnamefont {Derzsi}}, \bibinfo {author} {\bibfnamefont
  {P.}~\bibnamefont {Barone}}, \bibinfo {author} {\bibfnamefont
  {K.}~\bibnamefont {Wohlfeld}}, \bibinfo {author} {\bibfnamefont
  {J.}~\bibnamefont {Lorenzana}}, \ and\ \bibinfo {author} {\bibfnamefont
  {W.}~\bibnamefont {Grochala}},\ }\href {\doibase 10.1073/pnas.1812857116}
  {\bibfield  {journal} {\bibinfo  {journal} {Proceedings of the National
  Academy of Sciences}\ }\textbf {\bibinfo {volume} {116}},\ \bibinfo {pages}
  {1495} (\bibinfo {year} {2019})}\BibitemShut {NoStop}%
\bibitem [{\citenamefont {Hirayama}\ \emph {et~al.}(2020)\citenamefont
  {Hirayama}, \citenamefont {Tadano}, \citenamefont {Nomura},\ and\
  \citenamefont {Arita}}]{arita20prb}%
  \BibitemOpen
  \bibfield  {author} {\bibinfo {author} {\bibfnamefont {M.}~\bibnamefont
  {Hirayama}}, \bibinfo {author} {\bibfnamefont {T.}~\bibnamefont {Tadano}},
  \bibinfo {author} {\bibfnamefont {Y.}~\bibnamefont {Nomura}}, \ and\ \bibinfo
  {author} {\bibfnamefont {R.}~\bibnamefont {Arita}},\ }\href {\doibase
  10.1103/PhysRevB.101.075107} {\bibfield  {journal} {\bibinfo  {journal}
  {Phys. Rev. B}\ }\textbf {\bibinfo {volume} {101}},\ \bibinfo {pages}
  {075107} (\bibinfo {year} {2020})}\BibitemShut {NoStop}%
\bibitem [{\citenamefont {Bernardini}\ \emph {et~al.}(2020)\citenamefont
  {Bernardini}, \citenamefont {Olevano}, \citenamefont {Blase},\ and\
  \citenamefont {Cano}}]{cano20d}%
  \BibitemOpen
  \bibfield  {author} {\bibinfo {author} {\bibfnamefont {F.}~\bibnamefont
  {Bernardini}}, \bibinfo {author} {\bibfnamefont {V.}~\bibnamefont {Olevano}},
  \bibinfo {author} {\bibfnamefont {X.}~\bibnamefont {Blase}}, \ and\ \bibinfo
  {author} {\bibfnamefont {A.}~\bibnamefont {Cano}},\ }\href {\doibase
  10.1088/2515-7639/ab885d} {\bibfield  {journal} {\bibinfo  {journal} {J.
  Phys. Mater.}\ }\textbf {\bibinfo {volume} {3}},\ \bibinfo {pages} {035003}
  (\bibinfo {year} {2020})}\BibitemShut {NoStop}%
\bibitem [{\citenamefont {Wissel}\ \emph {et~al.}(2020)\citenamefont {Wissel},
  \citenamefont {Malik}, \citenamefont {Vasala}, \citenamefont {Plana-Ruiz},
  \citenamefont {Kolb}, \citenamefont {Slater}, \citenamefont {da~Silva},
  \citenamefont {Alff}, \citenamefont {Rohrer},\ and\ \citenamefont
  {Clemens}}]{clemens20}%
  \BibitemOpen
  \bibfield  {author} {\bibinfo {author} {\bibfnamefont {K.}~\bibnamefont
  {Wissel}}, \bibinfo {author} {\bibfnamefont {A.~M.}\ \bibnamefont {Malik}},
  \bibinfo {author} {\bibfnamefont {S.}~\bibnamefont {Vasala}}, \bibinfo
  {author} {\bibfnamefont {S.}~\bibnamefont {Plana-Ruiz}}, \bibinfo {author}
  {\bibfnamefont {U.}~\bibnamefont {Kolb}}, \bibinfo {author} {\bibfnamefont
  {P.~R.}\ \bibnamefont {Slater}}, \bibinfo {author} {\bibfnamefont
  {I.}~\bibnamefont {da~Silva}}, \bibinfo {author} {\bibfnamefont
  {L.}~\bibnamefont {Alff}}, \bibinfo {author} {\bibfnamefont {J.}~\bibnamefont
  {Rohrer}}, \ and\ \bibinfo {author} {\bibfnamefont {O.}~\bibnamefont
  {Clemens}},\ }\href {\doibase 10.1021/acs.chemmater.0c00193} {\bibfield
  {journal} {\bibinfo  {journal} {Chemistry of Materials}\ }\textbf {\bibinfo
  {volume} {32}},\ \bibinfo {pages} {3160} (\bibinfo {year}
  {2020})}\BibitemShut {NoStop}%
\bibitem [{\citenamefont {Botana}\ and\ \citenamefont
  {Norman}(2020)}]{botana20prx}%
  \BibitemOpen
  \bibfield  {author} {\bibinfo {author} {\bibfnamefont {A.~S.}\ \bibnamefont
  {Botana}}\ and\ \bibinfo {author} {\bibfnamefont {M.~R.}\ \bibnamefont
  {Norman}},\ }\href {\doibase 10.1103/PhysRevX.10.011024} {\bibfield
  {journal} {\bibinfo  {journal} {Phys. Rev. X}\ }\textbf {\bibinfo {volume}
  {10}},\ \bibinfo {pages} {011024} (\bibinfo {year} {2020})}\BibitemShut
  {NoStop}%
\bibitem [{\citenamefont {Olevano}\ \emph {et~al.}(2020)\citenamefont
  {Olevano}, \citenamefont {Bernardini}, \citenamefont {Blase},\ and\
  \citenamefont {Cano}}]{cano20b}%
  \BibitemOpen
  \bibfield  {author} {\bibinfo {author} {\bibfnamefont {V.}~\bibnamefont
  {Olevano}}, \bibinfo {author} {\bibfnamefont {F.}~\bibnamefont {Bernardini}},
  \bibinfo {author} {\bibfnamefont {X.}~\bibnamefont {Blase}}, \ and\ \bibinfo
  {author} {\bibfnamefont {A.}~\bibnamefont {Cano}},\ }\href {\doibase
  10.1103/PhysRevB.101.161102} {\bibfield  {journal} {\bibinfo  {journal}
  {Phys. Rev. B}\ }\textbf {\bibinfo {volume} {101}},\ \bibinfo {pages}
  {161102} (\bibinfo {year} {2020})}\BibitemShut {NoStop}%
\bibitem [{\citenamefont {{Lechermann}}()}]{lechermann20}%
  \BibitemOpen
  \bibfield  {author} {\bibinfo {author} {\bibfnamefont {F.}~\bibnamefont
  {{Lechermann}}},\ }\href@noop {} {\ }\Eprint
  {http://arxiv.org/abs/2012.09796} {arXiv:2012.09796} \BibitemShut {NoStop}%
\bibitem [{\citenamefont {Nica}\ \emph {et~al.}(2020)\citenamefont {Nica},
  \citenamefont {Krishna}, \citenamefont {Yu}, \citenamefont {Si},
  \citenamefont {Botana},\ and\ \citenamefont {Erten}}]{botana20trilayer}%
  \BibitemOpen
  \bibfield  {author} {\bibinfo {author} {\bibfnamefont {E.~M.}\ \bibnamefont
  {Nica}}, \bibinfo {author} {\bibfnamefont {J.}~\bibnamefont {Krishna}},
  \bibinfo {author} {\bibfnamefont {R.}~\bibnamefont {Yu}}, \bibinfo {author}
  {\bibfnamefont {Q.}~\bibnamefont {Si}}, \bibinfo {author} {\bibfnamefont
  {A.~S.}\ \bibnamefont {Botana}}, \ and\ \bibinfo {author} {\bibfnamefont
  {O.}~\bibnamefont {Erten}},\ }\href {\doibase 10.1103/PhysRevB.102.020504}
  {\bibfield  {journal} {\bibinfo  {journal} {Phys. Rev. B}\ }\textbf {\bibinfo
  {volume} {102}},\ \bibinfo {pages} {020504} (\bibinfo {year}
  {2020})}\BibitemShut {NoStop}%
\bibitem [{\citenamefont {Blaha}\ \emph {et~al.}()\citenamefont {Blaha},
  \citenamefont {Schwarz}, \citenamefont {Madsen}, \citenamefont {Kvasnicka},
  \citenamefont {Luitz}, \citenamefont {Laskowski}, \citenamefont {Tran},\ and\
  \citenamefont {Marks}}]{WIEN2k}%
  \BibitemOpen
  \bibfield  {author} {\bibinfo {author} {\bibfnamefont {P.}~\bibnamefont
  {Blaha}}, \bibinfo {author} {\bibfnamefont {K.}~\bibnamefont {Schwarz}},
  \bibinfo {author} {\bibfnamefont {G.}~\bibnamefont {Madsen}}, \bibinfo
  {author} {\bibfnamefont {D.}~\bibnamefont {Kvasnicka}}, \bibinfo {author}
  {\bibfnamefont {J.}~\bibnamefont {Luitz}}, \bibinfo {author} {\bibfnamefont
  {R.}~\bibnamefont {Laskowski}}, \bibinfo {author} {\bibfnamefont
  {F.}~\bibnamefont {Tran}}, \ and\ \bibinfo {author} {\bibfnamefont {L.~D.}\
  \bibnamefont {Marks}},\ }\href@noop {} {\bibinfo  {journal} {{W}{I}{E}{N}2k,
  An Augmented Plane Wave + Local Orbitals Program for Calculating Crystal
  Properties (Karlheinz Schwarz, Techn. Universität Wien, Austria), 2018. ISBN
  3-9501031-1-2}\ }\BibitemShut {NoStop}%
\bibitem [{\citenamefont {Chattopadhyay}\ \emph {et~al.}(1991)\citenamefont
  {Chattopadhyay}, \citenamefont {Brown},\ and\ \citenamefont
  {Köbler}}]{kobler91}%
  \BibitemOpen
\bibfield  {journal} {  }\bibfield  {author} {\bibinfo {author} {\bibfnamefont
  {T.}~\bibnamefont {Chattopadhyay}}, \bibinfo {author} {\bibfnamefont
  {P.}~\bibnamefont {Brown}}, \ and\ \bibinfo {author} {\bibfnamefont
  {U.}~\bibnamefont {Köbler}},\ }\href {\doibase
  https://doi.org/10.1016/0921-4534(91)90482-E} {\bibfield  {journal} {\bibinfo
   {journal} {Physica C: Superconductivity}\ }\textbf {\bibinfo {volume}
  {177}},\ \bibinfo {pages} {294} (\bibinfo {year} {1991})}\BibitemShut
  {NoStop}%
\bibitem [{\citenamefont {Perdew}\ \emph {et~al.}(1996)\citenamefont {Perdew},
  \citenamefont {Burke},\ and\ \citenamefont {Ernzerhof}}]{PBE}%
  \BibitemOpen
  \bibfield  {author} {\bibinfo {author} {\bibfnamefont {J.~P.}\ \bibnamefont
  {Perdew}}, \bibinfo {author} {\bibfnamefont {K.}~\bibnamefont {Burke}}, \
  and\ \bibinfo {author} {\bibfnamefont {M.}~\bibnamefont {Ernzerhof}},\ }\href
  {\doibase 10.1103/PhysRevLett.77.3865} {\bibfield  {journal} {\bibinfo
  {journal} {Phys. Rev. Lett.}\ }\textbf {\bibinfo {volume} {77}},\ \bibinfo
  {pages} {3865} (\bibinfo {year} {1996})}\BibitemShut {NoStop}%
\bibitem [{\citenamefont {Marzari}\ \emph {et~al.}(2012)\citenamefont
  {Marzari}, \citenamefont {Mostofi}, \citenamefont {Yates}, \citenamefont
  {Souza},\ and\ \citenamefont {Vanderbilt}}]{MLWF}%
  \BibitemOpen
  \bibfield  {author} {\bibinfo {author} {\bibfnamefont {N.}~\bibnamefont
  {Marzari}}, \bibinfo {author} {\bibfnamefont {A.~A.}\ \bibnamefont
  {Mostofi}}, \bibinfo {author} {\bibfnamefont {J.~R.}\ \bibnamefont {Yates}},
  \bibinfo {author} {\bibfnamefont {I.}~\bibnamefont {Souza}}, \ and\ \bibinfo
  {author} {\bibfnamefont {D.}~\bibnamefont {Vanderbilt}},\ }\href {\doibase
  10.1103/RevModPhys.84.1419} {\bibfield  {journal} {\bibinfo  {journal} {Rev.
  Mod. Phys.}\ }\textbf {\bibinfo {volume} {84}},\ \bibinfo {pages} {1419}
  (\bibinfo {year} {2012})}\BibitemShut {NoStop}%
\bibitem [{\citenamefont {Mostofi}\ \emph {et~al.}(2008)\citenamefont
  {Mostofi}, \citenamefont {Yates}, \citenamefont {Lee}, \citenamefont {Souza},
  \citenamefont {Vanderbilt},\ and\ \citenamefont {Marzari}}]{Wannier90}%
  \BibitemOpen
  \bibfield  {author} {\bibinfo {author} {\bibfnamefont {A.~A.}\ \bibnamefont
  {Mostofi}}, \bibinfo {author} {\bibfnamefont {J.~R.}\ \bibnamefont {Yates}},
  \bibinfo {author} {\bibfnamefont {Y.-S.}\ \bibnamefont {Lee}}, \bibinfo
  {author} {\bibfnamefont {I.}~\bibnamefont {Souza}}, \bibinfo {author}
  {\bibfnamefont {D.}~\bibnamefont {Vanderbilt}}, \ and\ \bibinfo {author}
  {\bibfnamefont {N.}~\bibnamefont {Marzari}},\ }\href {\doibase
  https://doi.org/10.1016/j.cpc.2007.11.016} {\bibfield  {journal} {\bibinfo
  {journal} {Computer Physics Communications}\ }\textbf {\bibinfo {volume}
  {178}},\ \bibinfo {pages} {685 } (\bibinfo {year} {2008})}\BibitemShut
  {NoStop}%
\bibitem [{\citenamefont {Kunes}\ \emph {et~al.}(2010)\citenamefont {Kunes},
  \citenamefont {Arita}, \citenamefont {Wissgott}, \citenamefont {Toschi},
  \citenamefont {Ikeda},\ and\ \citenamefont {Held}}]{Wien2wannier}%
  \BibitemOpen
  \bibfield  {author} {\bibinfo {author} {\bibfnamefont {J.}~\bibnamefont
  {Kunes}}, \bibinfo {author} {\bibfnamefont {R.}~\bibnamefont {Arita}},
  \bibinfo {author} {\bibfnamefont {P.}~\bibnamefont {Wissgott}}, \bibinfo
  {author} {\bibfnamefont {A.}~\bibnamefont {Toschi}}, \bibinfo {author}
  {\bibfnamefont {H.}~\bibnamefont {Ikeda}}, \ and\ \bibinfo {author}
  {\bibfnamefont {K.}~\bibnamefont {Held}},\ }\href {\doibase
  https://doi.org/10.1016/j.cpc.2010.08.005} {\bibfield  {journal} {\bibinfo
  {journal} {Computer Physics Communications}\ }\textbf {\bibinfo {volume}
  {181}},\ \bibinfo {pages} {1888 } (\bibinfo {year} {2010})}\BibitemShut
  {NoStop}%
\bibitem [{\citenamefont {Perdew}\ and\ \citenamefont {Zunger}(1981)}]{LDA}%
  \BibitemOpen
  \bibfield  {author} {\bibinfo {author} {\bibfnamefont {J.~P.}\ \bibnamefont
  {Perdew}}\ and\ \bibinfo {author} {\bibfnamefont {A.}~\bibnamefont
  {Zunger}},\ }\href {\doibase 10.1103/PhysRevB.23.5048} {\bibfield  {journal}
  {\bibinfo  {journal} {Phys. Rev. B}\ }\textbf {\bibinfo {volume} {23}},\
  \bibinfo {pages} {5048} (\bibinfo {year} {1981})}\BibitemShut {NoStop}%
\bibitem [{\citenamefont {{Bernardini}}\ and\ \citenamefont
  {{Cano}}(2020)}]{cano20c}%
  \BibitemOpen
  \bibfield  {author} {\bibinfo {author} {\bibfnamefont {F.}~\bibnamefont
  {{Bernardini}}}\ and\ \bibinfo {author} {\bibfnamefont {A.}~\bibnamefont
  {{Cano}}},\ }\href {\doibase 10.1088/2515-7639/ab9d0f} {\bibfield  {journal}
  {\bibinfo  {journal} {J. Phys. Mater.}\ }\textbf {\bibinfo {volume} {3}},\
  \bibinfo {eid} {03LT01} (\bibinfo {year} {2020})}\BibitemShut {NoStop}%
\bibitem [{\citenamefont {Pavarini}\ \emph {et~al.}(2001)\citenamefont
  {Pavarini}, \citenamefont {Dasgupta}, \citenamefont {Saha-Dasgupta},
  \citenamefont {Jepsen},\ and\ \citenamefont {Andersen}}]{andersen01-prl}%
  \BibitemOpen
  \bibfield  {author} {\bibinfo {author} {\bibfnamefont {E.}~\bibnamefont
  {Pavarini}}, \bibinfo {author} {\bibfnamefont {I.}~\bibnamefont {Dasgupta}},
  \bibinfo {author} {\bibfnamefont {T.}~\bibnamefont {Saha-Dasgupta}}, \bibinfo
  {author} {\bibfnamefont {O.}~\bibnamefont {Jepsen}}, \ and\ \bibinfo {author}
  {\bibfnamefont {O.~K.}\ \bibnamefont {Andersen}},\ }\href {\doibase
  10.1103/PhysRevLett.87.047003} {\bibfield  {journal} {\bibinfo  {journal}
  {Phys. Rev. Lett.}\ }\textbf {\bibinfo {volume} {87}},\ \bibinfo {pages}
  {047003} (\bibinfo {year} {2001})}\BibitemShut {NoStop}%
\bibitem [{\citenamefont {Sakakibara}\ \emph {et~al.}(2012)\citenamefont
  {Sakakibara}, \citenamefont {Usui}, \citenamefont {Kuroki}, \citenamefont
  {Arita},\ and\ \citenamefont {Aoki}}]{aoki12-prb}%
  \BibitemOpen
  \bibfield  {author} {\bibinfo {author} {\bibfnamefont {H.}~\bibnamefont
  {Sakakibara}}, \bibinfo {author} {\bibfnamefont {H.}~\bibnamefont {Usui}},
  \bibinfo {author} {\bibfnamefont {K.}~\bibnamefont {Kuroki}}, \bibinfo
  {author} {\bibfnamefont {R.}~\bibnamefont {Arita}}, \ and\ \bibinfo {author}
  {\bibfnamefont {H.}~\bibnamefont {Aoki}},\ }\href {\doibase
  10.1103/PhysRevB.85.064501} {\bibfield  {journal} {\bibinfo  {journal} {Phys.
  Rev. B}\ }\textbf {\bibinfo {volume} {85}},\ \bibinfo {pages} {064501}
  (\bibinfo {year} {2012})}\BibitemShut {NoStop}%
\bibitem [{\citenamefont {Jang}\ \emph {et~al.}(2016)\citenamefont {Jang},
  \citenamefont {Sakakibara}, \citenamefont {Kino}, \citenamefont {Kotani},
  \citenamefont {Kuroki},\ and\ \citenamefont {Han}}]{jang16sr}%
  \BibitemOpen
  \bibfield  {author} {\bibinfo {author} {\bibfnamefont {S.~W.}\ \bibnamefont
  {Jang}}, \bibinfo {author} {\bibfnamefont {H.}~\bibnamefont {Sakakibara}},
  \bibinfo {author} {\bibfnamefont {H.}~\bibnamefont {Kino}}, \bibinfo {author}
  {\bibfnamefont {T.}~\bibnamefont {Kotani}}, \bibinfo {author} {\bibfnamefont
  {K.}~\bibnamefont {Kuroki}}, \ and\ \bibinfo {author} {\bibfnamefont {M.~J.}\
  \bibnamefont {Han}},\ }\href@noop {} {\bibfield  {journal} {\bibinfo
  {journal} {Scientific reports}\ }\textbf {\bibinfo {volume} {6}},\ \bibinfo
  {pages} {1} (\bibinfo {year} {2016})}\BibitemShut {NoStop}%
\end{thebibliography}%

\end{document}